\documentclass[sigconf]{acmart}
\RequirePackage{rotating}
\RequirePackage{fix-cm}
\usepackage{amsmath}
\usepackage{ascmac}
\usepackage{color}
\usepackage{graphicx}
\usepackage[T1]{fontenc}
\usepackage{wrapfig}
\input{insbox}
\usepackage{graphicx}
\usepackage[misc,geometry]{ifsym}
\usepackage{adjustbox}
\usepackage{framed}
\usepackage[utf8]{inputenc}
\usepackage{booktabs}
\usepackage{subcaption}
\usepackage{tcolorbox}
\usepackage{amsmath}
\usepackage{paralist}
\usepackage{multirow}
\usepackage{xspace}
\usepackage{color}
\usepackage{xcolor}
\usepackage[graphicx]{realboxes}
\usepackage{ifthen}
\usepackage{url}
\usepackage{fancybox}
\usepackage{enumitem}
\usepackage{listings}
\usepackage{balance}
\usepackage{ifthen}
\usepackage{graphicx}
\usepackage[linesnumbered,boxruled]{algorithm2e}
\usepackage{tablefootnote}
\usepackage{comment}
\usepackage{float}
\usepackage{algorithmic}
\usepackage{dirtytalk}
\usepackage{soul}
\usepackage{tcolorbox}
\usepackage{colortbl}
\usepackage[normalem]{ulem}
\usepackage{booktabs,tabularx,enumitem}
\usepackage{xurl}

\copyrightyear{2026}
\acmYear{2026}
\setcopyright{cc}
\setcctype{by}
\acmConference[MSR '26]{23rd International Conference on Mining Software Repositories}{April 13--14, 2026}{Rio de Janeiro, Brazil}
\acmBooktitle{23rd International Conference on Mining Software Repositories (MSR '26), April 13--14, 2026, Rio de Janeiro, Brazil}
\acmPrice{}
\acmDOI{10.1145/3793302.3793336}
\acmISBN{979-8-4007-2474-9/2026/04}

\newlist{rqlist}{description}{1}
\setlist[rqlist]{%
  font=\bfseries,      
  style=nextline,      
  labelsep=0.6em,      
  leftmargin=!,        
  widest={(RQ3)},      
  itemsep=.6ex, topsep=.6ex
}

\def\summaryblock#1#2{
\begin{oframed}
\noindent \textbf{#1:}
#2
\end{oframed}
}

\newcommand{\Motivation}{\textit{\underline{Motivation:}} }
\newcommand{\Results}{\textit{\underline{Results:}} }

\input{insbox}

\definecolor{mygray}{gray}{0.6}

\newlength\WIDTHOFBAR
\DeclareGraphicsExtensions{.pdf,.jpeg,.png}
\newcommand{\RQone}{To what extent can the pipeline successfully generate traceability links for declined proposals?}
\newcommand{\RQtwo}{Why does the pipeline fail to generate traceability links?}

\begin{document}

\title{Toward Linking Declined Proposals and Source Code:\\An
Exploratory Study on the Go Repository}


\author{Sota Nakashima}
\affiliation{%
  \institution{Kyushu University}
  \country{Japan}
	}
\email{nakashima@posl.ait.kyushu-u.ac.jp}

\author{Masanari Kondo}
\affiliation{%
  \institution{Kyushu University}
  \country{Japan}
	}
\email{kondo@ait.kyushu-u.ac.jp}

\author{Mahmoud Alfadel}
\affiliation{%
  \institution{University of Calgary}
  \country{Canada}
	}
\email{mahmoud.alfadel@ucalgary.ca}

\author{Aly Ahmad}
\affiliation{%
  \institution{University of Calgary} 
  \country{Canada}
	}
\email{aly.ahmad@ucalgary.ca}

\author{Toshihiro Nakae}
\affiliation{%
  \institution{DENSO CORPORATION} 
  \country{Japan}
	}
\email{toshihiro.nakae.j8z@jp.denso.com}

\author{Hidenori Matsuzaki}
\affiliation{%
  \institution{DENSO CORPORATION}
  \country{Japan}
	}
 \email{hidenori.matsuzaki.j4f@jp.denso.com}

\author{Yasutaka Kamei}
\affiliation{%
  \institution{Kyushu University}
  \country{Japan}
	}
\email{kamei@ait.kyushu-u.ac.jp}



\begin{abstract}
Traceability links are key information sources for software developers, connecting software artifacts.
Such links play an important role, particularly between contribution artifacts and their corresponding source code.
Through these links, developers can trace the discussions in contributions and uncover design rationales, constraints, and security concerns.
Previous studies have mainly examined accepted contributions, while those declined after discussion have been overlooked.
Declined-contribution discussions capture valuable design rationale and implicit decision criteria, revealing why features are accepted or rejected. 
Our prior work also shows developers often revisit and resubmit declined contributions, making traceability to them useful.

In this study, we present the first attempt to establish traceability links between declined contributions and related source code.
We propose a linking approach and conduct an empirical analysis of the generated links to discuss the factors that affect link generation.
As our dataset, we use \textit{proposals} from the official Go repository, which are GitHub issues used to propose new features or language changes.
To link declined proposals to source code, we design an LLM-driven pipeline.
Our results show that the pipeline selected the correct granularity for each declined proposal with an accuracy of 0.836, and generated correct links at that granularity with a mean precision of 0.643. 
To clarify the challenges of linking declined proposals, we conduct a failure analysis of instances where the pipeline failed to generate links. In these cases, discussions were often redundant and lacked concrete information (e.g., details on how the feature should be implemented).
\end{abstract}



\maketitle

\vspace{-2mm}
\section{Introduction}
\label{sec:intro}

Software traceability (i.e., the establishment of explicit links between software artifacts such as issue reports, design documents and source code changes) is a cornerstone of effective software development and maintenance. 
These traceability links support a wide range of tasks such as bug localization, defect prediction, and code comprehension. In both industrial~\cite{cleland2012software, cleland2014trends} and open-source settings~\cite{aung2020literature, nguyen2022hermes, naslavsky2007using, panis2010successful}, traceability links serve as rich knowledge sources that help developers understand past decisions, align artifacts, and reduce latent maintenance costs.

\begin{figure}[t]
    \begin{center}
    \includegraphics[width=\linewidth]{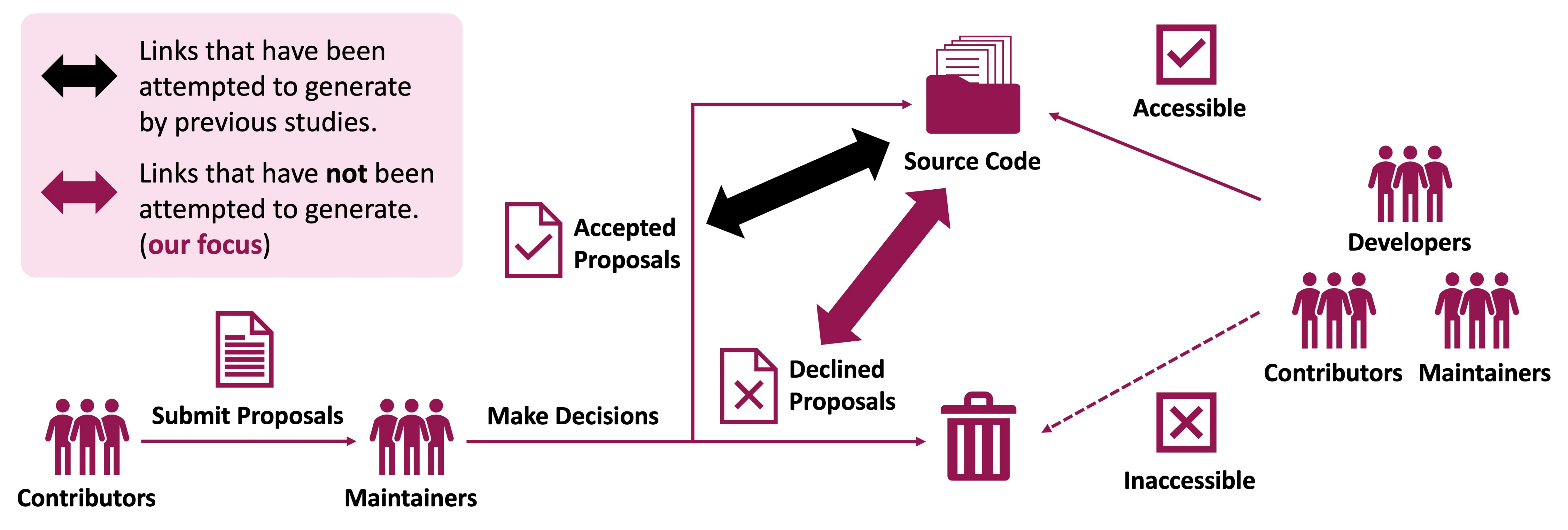}
    \vspace{-6pt}
    \caption{Development workflow of proposals}
    \label{fig:background}
    \end{center}
\end{figure}

In the open-source ecosystem, an especially important category of traceability link is that between submitted contributions (for new features, bug fixes or refactorings; hereafter referred to as \textit{proposals}) and the source code changes that implement them. Figure~\ref{fig:background} illustrates how contributors propose changes, maintainers review them, and accepted proposals lead to source code evolution. Proposals (and their associated discussions) capture essential information about rationale, technical constraints, and community consensus.
Prior work~\cite{wang2025mplinker, lan2023btlink} has leveraged such links to support automated tooling and analytics.  


While previous studies focused on accepted proposals that were implemented in the source code, \textit{declined} proposals that were not implemented have received little attention.
Despite their lack of implementation, these declined proposals often contain valuable insights, such as design alternatives, failed decisions, or rejected trade-offs, that are highly relevant to future contributors, maintainers and researchers alike. 
In fact, our prior work shows that 14.7\% of declined proposals were later resubmitted as new proposals~\cite{kondo2025empirical}, suggesting that developers do revisit prior declined discussions and rationales when refining ideas.

Therefore, in this paper, we focus on \textit{linking declined proposals to source code}.
We define a novel linking task, propose a pipeline driven by large language models (LLMs), and carry out an empirical study on the proposals of the official Go repository~\cite{go_proposal_readme}, which offers a rigorous and well-documented review process~\cite{kondo2025empirical}. 
The initial linking approach consists of two steps: (1) determining the appropriate granularity of source code to link for each proposal (e.g., directory, file, or function level), and (2) generating links at that granularity. We use LLMs to implement this approach, given their demonstrated effectiveness in various software engineering tasks~\cite{wei2024magicoder, yuan2023evaluating, jin2023inferfix}.
Finally, we evaluate how well our pipeline automatically selects the appropriate code granularity and generates candidate links, and then analyze failure cases to surface patterns and challenges that are unique to this scenario.

Specifically, we addressed the following research questions (RQs):

\begin{rqlist}
  \item[(RQ1) \RQone]
  \Motivation
Prior work has focused on linking accepted proposals to source code, but it remains unclear how effectively declined proposals can also be linked. Assessing this capability reveals the potential of automated methods to recover knowledge from unimplemented discussions.\\
  \Results
  Our proposed pipeline achieved an average accuracy of 0.836 in selecting an appropriate source code granularity and a precision of 0.643 in correctly generating links to the relevant targets at that granularity.

  \item[(RQ2) \RQtwo]
  \Motivation
 Understanding the causes of link-generation failures can expose the limitations of current techniques and guide future improvements.\\
  \Results
  The performance of the pipeline heavily relies on the discussion content of declined proposals. In particular, the pipeline often failed when the discussions described what to implement but did not include concrete information on how to implement it.
\end{rqlist}

In summary, this paper makes the following contributions:
\begin{itemize}\setlength{\itemsep}{0pt}
\item \textbf{Novel task:} We present the first study to establish traceability links between declined proposals and source code, making the valuable knowledge embedded in unimplemented proposals accessible to developers and researchers.
  \item \textbf{LLM-driven approach:} We design and evaluate an LLM-driven pipeline for traceability link recovery using proposals from the official Go language repository. Both the pipeline and our experimental results are publicly available in the replication package~\cite{naka2025replication}.
  \item \textbf{Failure analysis:} We conduct a detailed failure analysis to identify characteristics of declined proposals that influence linking performance, providing insights to guide future improvements in automated traceability.
\end{itemize}

\begin{figure*}[t]
    \begin{center}
    \includegraphics[width=\linewidth]{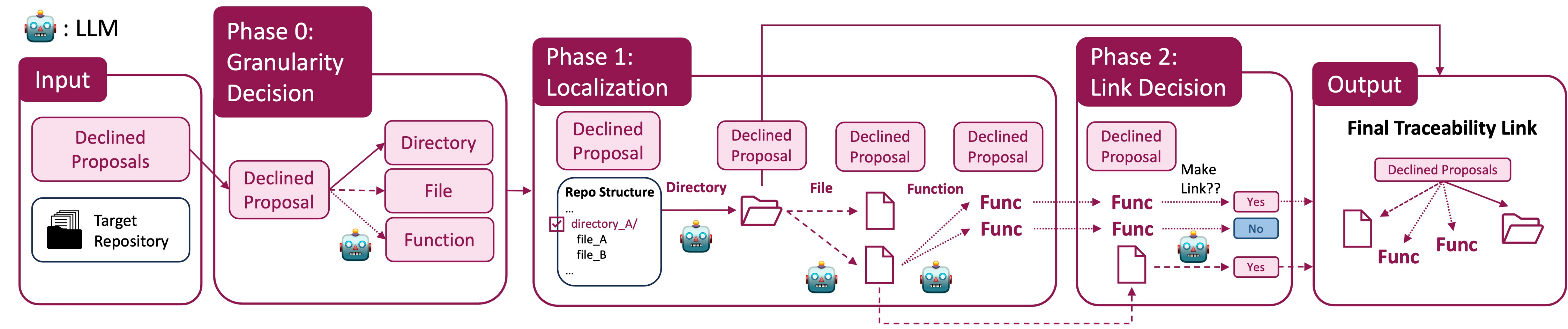}
    \vspace{-6pt}
    \caption{LLM-driven pipeline for generating granularity-aware traceability links}
    \label{fig:workflow}
    \end{center}
\end{figure*}


\section{Related Work}
\label{sec:related}
In this section, we discuss the related work and position our study within the literature.

\subsection{Declined Contributions}
Declined contributions to OSS projects (e.g., pull requests and issues that are closed without merged) have been examined to improve the development process and support contributors and maintainers. These studies seek to explain when and why such contributions fail to be merged, highlighting social, organizational, and coordination factors alongside technical concerns.

\noindent
\textbf{Rejected PRs.}
Several studies examine why pull requests (PRs) get rejected~\cite{gousios2014exploratory, steinmacher2018almost, li2021you, khatoonabadi2023wasted}.
Gousios et al.~\cite{gousios2014exploratory} analyzed 350 PRs that were closed without merged and found that typical challenges of distributed collaboration, especially coordination and communication, were more influential than code defects.
Steinmacher et al.~\cite{steinmacher2018almost} complemented this with a survey, noting duplication and misalignment with project goals as common reasons teams close PRs.
Li et al.~\cite{li2021you} manually analyzed 321 abandoned PRs from five OSS projects and surveyed 619 contributors and 91 integrators, identifying 12 common reasons.
However, it was not yet clear how PR, contributor, process, and project-level factors affected abandonment likelihood.
To address this gap, Khatoonabadi et al.~\cite{khatoonabadi2023wasted} studied a larger dataset of mature GitHub projects using quantitative and qualitative analyses.


\noindent
\textbf{Declined Proposals.}
Kondo et al.~\cite{kondo2025empirical} investigated declined proposals in the Go project to understand their declined reasons. They provided a taxonomy of decline reasons based on their manual coding. Also, they evaluated an LLM-based approach to predict whether a proposal would be accepted or declined.
The declined proposals sometimes involve design-level decisions and are especially important to maintainers because, once accepted, such design-level changes impose long-term maintenance obligations on the project~\cite{kondo2025empirical}.
Therefore, making the discussion, specifically the rationale behind declined proposals, readily accessible to future contributors is crucial.

Declined contributions typically do not appear in the codebase, making it difficult for contributors to access or reuse the knowledge contained in such declined contributions (Figure~\ref{fig:background}).
Prior studies~\cite{gousios2014exploratory, steinmacher2018almost, li2021you, khatoonabadi2023wasted, kondo2025empirical} have mainly analyzed the reasons for decline.
In contrast to these studies, our work proposes a new approach to establish traceability links between declined contributions and relevant source code.
This approach enables developers to directly access the knowledge of declined contributions from the codebase.



\vspace{-2mm}
\subsection{Traceability Link Recovery}
Traceability links in software engineering refer to the semantic relationships among various software artifacts, including requirements, design documents, and source code~\cite{cleland2012software, cleland2014trends, antoniol2002recovering}.
By capturing these relationships, traceability links provide additional contextual information that enhances the accuracy and effectiveness of various software engineering tasks, such as change impact analysis~\cite{aung2020literature}, vulnerability-fixing commit identification~\cite{nguyen2022hermes}, selective regression testing~\cite{naslavsky2007using}, and project management~\cite{panis2010successful}.
In practice, traceability links have often been maintained manually, which can lead to incomplete, inaccurate, non-scalable, and inconsistent records~\cite{cleland2012software, rodriguez2021leveraging}.
To address these limitations, a variety of automated traceability link recovery (TLR) techniques have been proposed.

A widely studied approach to automated TLR is based on information retrieval (IR) methods. These IR-based methods calculated text similarity by constructing retrieval models (e.g, Vector Space Model~\cite{hayes2006advancing}, Latent Semantic Indexing~\cite{de2004enhancing, rempel2013towards}, and Latent Dirichelet Allocation~\cite{dekhtyar2007technique, asuncion2010software}). However, these methods could not effectively uncover the semantic information between two artifacts~\cite{borg2014recovering}.

As an alternative to IR-based techniques, machine learning (ML) approaches have also been widely explored for TLR. The most common strategy is to train classifiers that predict whether a pair of artifacts should be linked based on their textual representations (e.g., FRLink~\cite{sun2017frlink}, Link Classifier~\cite{rath2018traceability}). However, such ML-based methods heavily rely on labeled training data and also struggle to capture the semantic relationships between artifacts since their semantic relationships are created by human. 
To enhance these ML approaches, deep learning (DL) techniques~\cite{xie2019deeplink, ruan2019deeplink} have been introduced to automatically extract semantic features and perform classification without manual feature engineering. Nevertheless, DL-based models still depend on large, high-quality datasets.


Large language models (LLMs) have delivered strong gains across software engineering tasks such as code generation~\cite{wei2024magicoder}, code summarization~\cite{yuan2023evaluating}, and program repair~\cite{jin2023inferfix}.
LLMs have also been shown to improve the performance of TLR. Because they are pre-trained on large-scale corpora, LLM-based approaches do not require task-specific labeled data.
For instance, Rodriguez et~al.~\cite{rodriguez2023prompts} explored prompt engineering to enhance TLR performance using LLMs.
Fuch{\ss} et~al.~combined LLMs with retrieval-augmented generation (RAG)~\cite{lewis2020retrieval} to incorporate project-relevant knowledge during inference, as demonstrated in LiSSA~\cite{fuchss2025lissa}.

\section{LLM-Driven Pipeline}
\label{sec:method}
In this section, we describe the LLM-driven pipeline we constructed to establish traceability links between declined proposals and the source code.


\subsection{Overview}
Figure~\ref{fig:workflow} illustrates the overall workflow of the pipeline. The process is informed by a state-of-the-art TLR method using retrieval augmented generation (RAG)~\cite{fuchss2025lissa}.
In the RAG-based method, embeddings are used to compute similarity between documents and source code to identify the code relevant to a document (\textit{Localization}). The document–code pairs are then provided to an LLM, which determines whether a link should be created (\textit{Link Decision}).
We extended the prior TLR method based on the features of declined proposals.
First, we added Phase~0 to determine the appropriate granularity of each proposal (\textit{Granularity Decision}) for generating granularity-aware traceability links.
Declined proposals vary widely in scope, ranging from abstract, design-level discussions to concrete suggestions such as modifying a specific function. Therefore, it is necessary to first determine the granularity of the discussion in order to decide the appropriate granularity of the code to be linked.
Second, we replaced the existing embedding-based localization with an LLM-based localization.
Purely embedding-based retrieval considers only similarity between proposals and source code. However,  the discussed code in the declined proposal may be absent, so highly similar artifacts are not necessarily relevant. 
Using an LLM enables context-aware localization that reasons over the discourse in declined proposals and repository structure.

\begin{figure}[t]
  \centering
  \includegraphics[width=\linewidth]{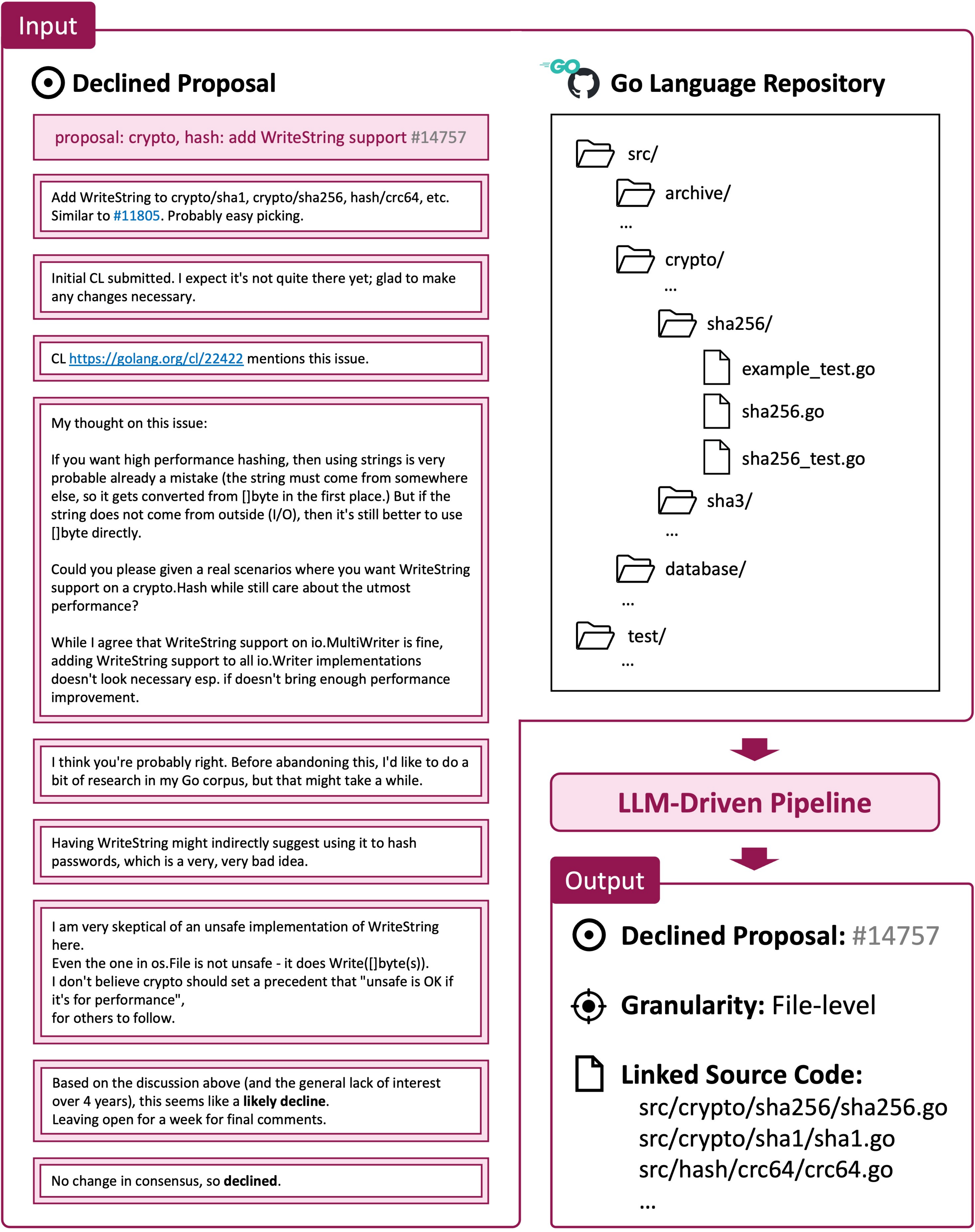}
  \caption[Input–output example from golang/go\#14757]%
  {Input–output example retrieved from issues/14757 in golang/go~\cite{issue_14757}}
  \label{fig:example}
\end{figure}

\noindent
\textbf{Input and Output of the Pipleine.}
The input comprises all discussions from the target declined proposals, along with repository information. The latter refers to the contextual data supplied to the LLM to support reasoning about the repository structure and its code artifacts; the specific types of context are detailed in subsequent subsections. 
The output is a set of traceability links connecting each declined proposal to related code artifacts at varying granularities. For each proposal, the pipeline first determines the most appropriate granularity and then establishes links to the corresponding code elements at that level.
An example of the actual input and output is shown in Figure~\ref{fig:example}.

\newpage

\vspace{-2mm}

\vspace{-2mm}

\subsection{Granularity Decision (Phase~0)}
In Phase~0, we determine the appropriate granularity of the source code linked to the given proposal. 
As shown in Figure~\ref{fig:workflow}, the LLM is provided only with the declined proposal and outputs an appropriate choice from various granularities. 
This granularity then dictates how the subsequent phases (Phase~1 and Phase~2) are carried out.

\begin{figure}[t]
    \begin{center}
    \includegraphics[width=.98\linewidth]{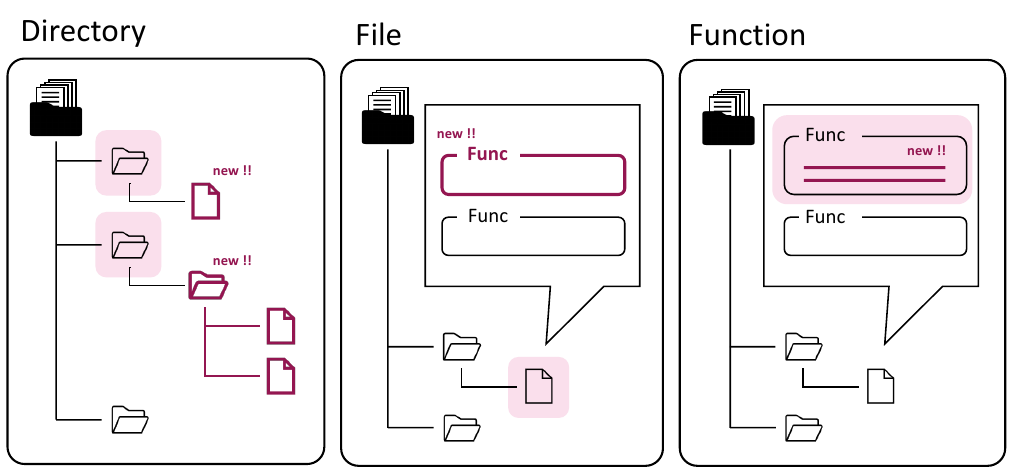}
    \caption{Three levels of granularity for declined proposals}
    \label{fig:granularity}
    \end{center}
\end{figure}

We define three levels of granularity for traceability links, as illustrated in Figure~\ref{fig:granularity}.
We describe each granularity level below.

\vspace{1mm}
\noindent
\textbf{Directory-level granularity.}
A proposal is classified at this level when it explicitly involves creating, removing, renaming, or moving files or directories (packages), thereby altering the package structure. Examples include adding new source files or resources, creating new packages, splitting code into separate files, or moving or renaming existing packages or files

\noindent
\textbf{File-level granularity.}
A proposal is classified at this level when all required edits are confined to existing files, even if the changes span multiple files or packages, without altering the file structure. Examples include adding new functions, methods, types, or constants within existing files.

\noindent
\textbf{Function-level granularity.}
A proposal is classified at this level when the change is confined to modifying the internal logic or documentation of existing functions or methods, without introducing new top-level declarations or requiring edits outside these functions or methods. This level applies when only statements, control flow, or function parameters and return values are modified, leaving the rest of the codebase unaffected. In this study, functions and methods refer to regular ones with executable bodies; built-in functions and interfaces are excluded.


\vspace{-2mm}
\subsection{Localization (Phase~1)}

In this phase, the LLM localizes the source code that relates to each declined proposal.
Our approach is inspired by Agentless \cite{xia2024agentless}, a fault-localization technique that identifies suspicious code locations (e.g., statements or methods) associated with bugs.
Agentless \cite{xia2024agentless} overcomes the limited context window of LLMs by applying a simplified hierarchical process.
Following this idea, we designed a three-step localization procedure, each step corresponding to one granularity level.

\vspace{-2mm}
\subsubsection{Directory-level Localization}
Following Agentless~\cite{xia2024agentless}, we represent the repository structure in a compact \textit{tree-like format} similar to the Linux \texttt{tree} command (Figure~\ref{fig:workflow}).
Instead of supplying the complete file contents of the repository, we build a compact map that starts at the root directory and lists folder and file names.
Entries at the same depth are aligned vertically, and subdirectories are indented.
We provide this representation to the LLM together with the full text of the declined proposal.
The LLM is prompted to identify directories relevant to the given proposal.
The output is a set of candidate directories, which are passed to subsequent steps.


\vspace{-2mm}
\subsubsection{File-level Localization}
Using the same inputs (repository structure and proposal text), the LLM is next prompted to identify specific files related to the proposal within the localized directories.
The model outputs a set of candidate files, which serve as inputs for the following stage.

\vspace{-2mm}
\subsubsection{Function-level Localization}
After obtaining the list of relevant files, the LLM localizes the functions or methods associated with the proposal.
Following Agentless~\cite{xia2024agentless}, we compress each file into a representation that includes only function and method \textit{signatures}, as illustrated below:
\begin{itemize}\setlength{\itemsep}{0pt}
  \item Functions: \texttt{func FuncName(...) \{ ... \}}
  \item Methods: \texttt{func (receiver) MethodName(...) \{ ... \}}
\end{itemize}
This format focuses solely on callable entities. 
Compared to providing entire file contents, this compact representation is far more efficient and cost-effective, especially when files contain thousands of lines.
Each file is processed independently in a separate prompt, enabling the LLM to focus on the signatures and identify the relevant functions and methods effectively.
The output is a list of function or method names deemed relevant to the proposal.

\vspace{-1mm}
\paragraph{\textbf{Granularity-aware Execution.}}
The granularity determined in Phase~0 dictates which localization steps are executed and how their outputs are propagated to subsequent phases (Figure~\ref{fig:workflow}).
For directory-level proposals, only directory-level localization is performed, and the resulting directories are directly used to generate final traceability links without proceeding to Phase~2.
For file-level proposals, both directory- and file-level localization are executed sequentially, and the identified files are passed to Phase~2 for link decision.
For function-level proposals, all three localization steps are executed in sequence, and the identified functions are passed to Phase~2.


\vspace{-2mm}
\subsection{Link Decision (Phase~2)}
In this phase, the LLM determines whether each localized source code element is relevant to the given declined proposal.
Building on prior work in TLR~\cite{fuchss2025lissa, rodriguez2023prompts}, we formulate this step as a binary classification task that establishes final traceability links.

\vspace{-1mm}
\subsubsection{Input and Task Definition}
The input to Phase~2 is a set of candidate pairs, each consisting of a declined proposal and one of the source code elements localized in Phase~1 (i.e., directories, files, or functions, depending on the granularity determined in Phase~0).
The task of the LLM is to judge, for each pair, whether the source code element is relevant to the given proposal.

\vspace{-1mm}
\subsubsection{Prompt Design and Output}
We prompt the LLM with the complete text of the declined proposal and the corresponding source code element, asking it to output a binary decision:
\begin{quote}
\texttt{Is this source code element relevant to the given proposal? (Yes / No)}
\end{quote}
The LLM then outputs either \texttt{Yes} or \texttt{No} to indicate relevance.
This design offers interpretability and avoids the need for arbitrary similarity thresholds often used in embedding-based retrieval methods.

\vspace{-2mm}
\section{Experiment setup}
\label{sec:research}
This section describes the studied project, selected models, evaluation metrics, and how we developed the prompts and evaluation criteria used in our pipeline.


\vspace{-2mm}
\subsection{Studied Project}
Prior work~\cite{kondo2025empirical} collected proposals from the official Go proposal process and released labels indicating whether each proposal was accepted or declined (337 accepted, 448 declined).
In this study, we use that curated proposal set together with the same Go codebase.\footnote{We analyzed the official Go repository pinned to commit \texttt{a8e99ab19cbf8568cb452b899d0ed3f0d65848c5} (SHA-1; retrieved on May 30, 2025).}
Our analysis focuses on \texttt{.go} source files. 
At this commit, the repository contains 1{,}468 directories, 10{,}605 files, and 85{,}800 functions/methods.

\vspace{-2mm}
\subsection{Selected Models}
The goal of this study is to define a pipeline for generating traceability links between declined proposals and source code and to evaluate that pipeline; benchmarking multiple LLMs is out of scope.
We therefore selected 
DeepSeek-V3.1 (non-thinking mode)~\cite{deepseek_v3_1_release} as the LLM for all experiments, with the temperature set to 0.0.
We find DeepSeek~\cite{liu2024deepseek, deepseek_v3_1_release} to be a cost-effective, competitive model for our setting; The pipeline processed each proposal in 43.7 seconds at a cost of \$0.115 on average.




\vspace{-2mm}
\subsection{Evaluation Metrics}
To evaluate the pipeline, we use common traceability metrics~\cite{hayes2006advancing, cleland2012software, fuchss2025lissa}: precision, recall, and F1 score.
We also assess whether the pipeline selects the correct code granularity for each proposal.

Each proposal $i$ is evaluated using four metrics, then aggregated by granularity and overall.
Let $g_i \in \{\textit{Directory}, \textit{File}, \textit{Function}\}$ denote the predicted granularity and $g_i^{*}$ the ground truth.
Let $P_i$ be the set of generated links and $G_i$ the ground-truth set.
We count (i) \emph{true positives} as links in both $P_i$ and $G_i$, (ii) \emph{false positives} as links in $P_i$ only, and (iii) \emph{false negatives} as links in $G_i$ only:
\[
\mathrm{TP}_i = |P_i \cap G_i|,\quad
\mathrm{FP}_i = |P_i \setminus G_i|,\quad
\mathrm{FN}_i = |G_i \setminus P_i|
\]

\begin{itemize}
  \item \textbf{Granularity Accuracy}:
  \[
  \mathrm{GA}_i = \mathbf{1}\!\left(g_i = g_i^{*}\right)\in\{0,1\}
  \]

  \item \textbf{Precision}:
  \[
  \mathrm{Precision}_i = \frac{\mathrm{TP}_i}{|P_i|}
  \]

  \item \textbf{Recall}:
  \[
  \mathrm{Recall}_i = \frac{\mathrm{TP}_i}{|G_i|}
  \]

  \item \textbf{F1 score}:
  \[
  \mathrm{F1}_i = \frac{2\,\mathrm{Precision}_i\,\mathrm{Recall}_i}{\mathrm{Precision}_i+\mathrm{Recall}_i}
  \]
\end{itemize}

\noindent
\textbf{Aggregation.}
For each per-proposal metric $M_i\in\{\mathrm{GA}_i,\mathrm{Prec}_i,\mathrm{Rec}_i,\mathrm{F1}_i\}$, we reported macro averages by granularity and overall.
For by-granularity averages, for each $b\in\{\textit{Directory},\textit{File},\textit{Function}\}$ we computed:
\[
\overline{M}^{\,\text{macro}}(b) =
\frac{\sum_i \mathbf{1}\!\left(g_i^{*}=b\right) M_i}{\sum_i \mathbf{1}\!\left(g_i^{*}=b\right)}
\]
The overall macro average is:
\[
\overline{M}^{\,\text{macro}} = \frac{\sum_i M_i}{\sum_i 1}
\]

\subsection{Prompts and Criteria Development}
\label{subsec:codebook_prompt}
We define the task of linking declined proposals to source code and design an LLM-based pipeline to address it.
Each phase of the pipeline uses a prompt tailored to its specific goal. 
For example, in Phase~1 (\textit{Granularity Decision}), the prompt instructs the LLM to determine the most suitable code granularity for each proposal.

Evaluating the pipeline requires criteria that account for the unique characteristics of declined proposals.
First, declined proposals often refer to code at different levels of granularity (e.g., directory, file, or function).  
Second, because these proposals were rejected, the code they discuss is typically not implemented in the repository.  
As a result, we cannot rely on automated matching between proposals and existing code.  
To evaluate the pipeline fairly, we developed two human-defined criteria: \textit{Granularity Criteria}, which guide the selection of the appropriate granularity for each proposal, and \textit{Link Correctness Criteria}, which guide the assessment of whether a generated link is reasonable given the proposal’s discussion.  
We summarize both below.

\newpage

\begin{itemize}\setlength{\itemsep}{0pt}
  \item \textbf{Granularity Criteria}:  
  define how to decide whether a proposal refers to a directory, file, or function level.  
  These criteria are used to construct the ground-truth granularity labels and to measure the pipeline’s granularity accuracy.
  
  \item \textbf{Link Correctness Criteria}:  
  specify how to judge whether a generated link correctly corresponds to the code discussed in a proposal (\textit{yes}/\textit{no}).  
  These criteria are used to assess link correctness and compute precision.
\end{itemize}

\noindent
The details are provided in our replication package~\cite{naka2025replication}.

The prompts and criteria were developed collaboratively by two authors through two steps: \textit{Initialization} and \textit{Validation}.

\noindent
\textbf{Initialization.}
The first author ran 20 declined proposals through the pipeline and refined the prompts and criteria based on the outputs.
The code categories used in the criteria (e.g., \textit{directory}, \textit{file}, and \textit{function} in the granularity criteria) were predefined.  
The goal at this stage was to establish clear decision boundaries between categories, not to introduce new ones.

\noindent
\textbf{Validation.}
To verify the consistency and reliability of the prompts and criteria, two authors (first and fourth) independently labeled proposals using both sets of criteria.  
Treating 20 proposals as one round, we computed Cohen’s $\kappa$ for both labeling tasks (granularity and link correctness) to assess inter-rater agreement, and we evaluated the prompts using granularity accuracy and precision.  
Thresholds and results are summarized in Table~\ref{tab:validation}.  
We adopted $\kappa \ge 0.61$ as the threshold for substantial agreement~\cite{viera2005understanding}.  
Although the first round met this threshold, we conducted one additional round to stabilize results.  
The validated prompts and criteria are available in our replication package~\cite{naka2025replication}.

In total, 20 declined proposals were used for \textit{Initialization} and 40 for \textit{Validation}, leaving 388 for the final evaluation.
This process mirrors the standard train–validation–test setup and follows a human-in-the-loop prompt engineering approach~\cite{shah2024prompt}.


\begin{table}[t!]
  \centering
  \caption{Thresholds and results for inter-rater agreement and performance}
  \vspace{-6pt}
  \label{tab:validation}
  \footnotesize 
  \setlength{\tabcolsep}{4pt} 
  \renewcommand{\arraystretch}{1.05} 
  \begin{tabular*}{\columnwidth}{@{\extracolsep{\fill}}lcccc@{}}
    \toprule
    & \multicolumn{2}{c}{Granularity} & \multicolumn{2}{c}{Link Correctness} \\
    \cmidrule(lr){2-3} \cmidrule(lr){4-5}
    & Cohen's $\kappa$ & Accuracy & Cohen's $\kappa$ & Precision \\
    \midrule
    Criteria & 0.610 & 0.750 & 0.610 & 0.750 \\
    1st & 0.632 & 0.929 & 0.714 & 0.902 \\
    2nd & 1.000 & 0.944 & 0.639 & 0.775 \\
    \bottomrule
  \end{tabular*}
  \vspace{-2mm}
\end{table}

\begin{figure*}[t]
    \centering
    \begin{subfigure}{0.33\textwidth}
        \centering
        \includegraphics[width=\linewidth]{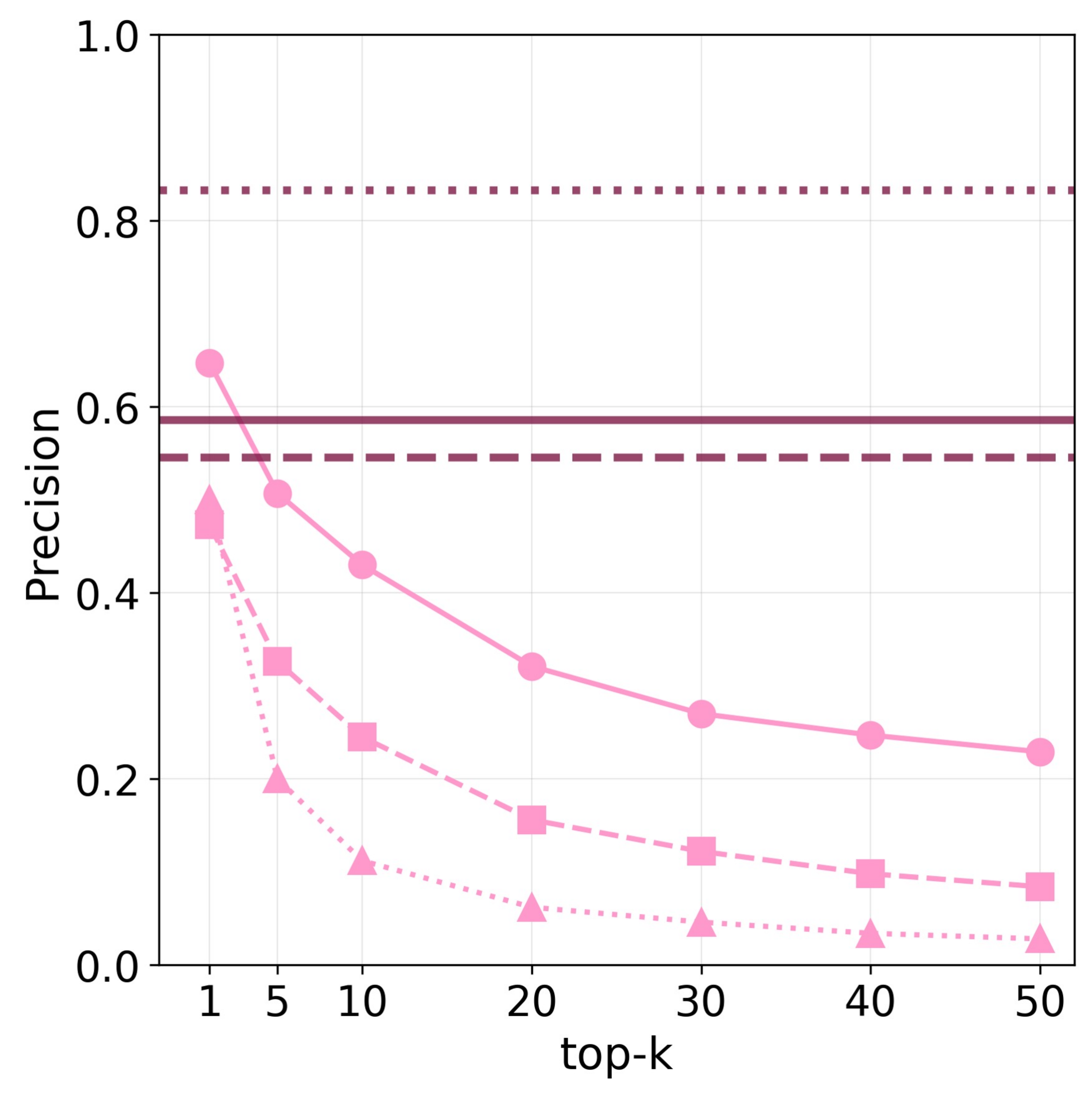}
        \vspace{-6pt}
        \caption{Precision}
        \label{fig:rq1_precision}
    \end{subfigure}\hfill
    \begin{subfigure}{0.33\textwidth}
        \centering
        \includegraphics[width=\linewidth]{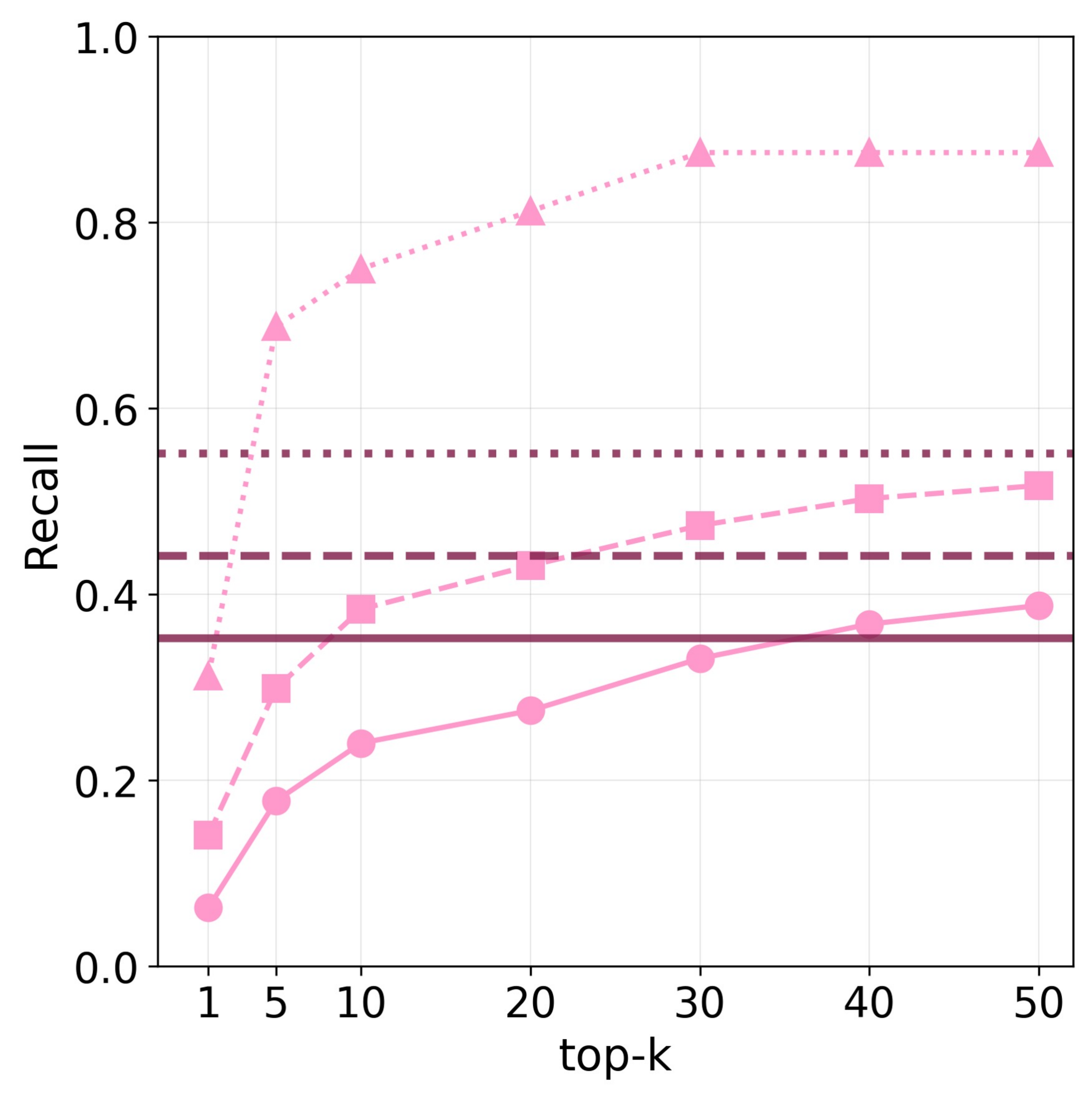}
        \vspace{-6pt}
        \caption{Recall}
        \label{fig:rq1_recall}
    \end{subfigure}\hfill
    \begin{subfigure}{0.33\textwidth}
        \centering
        \includegraphics[width=\linewidth]{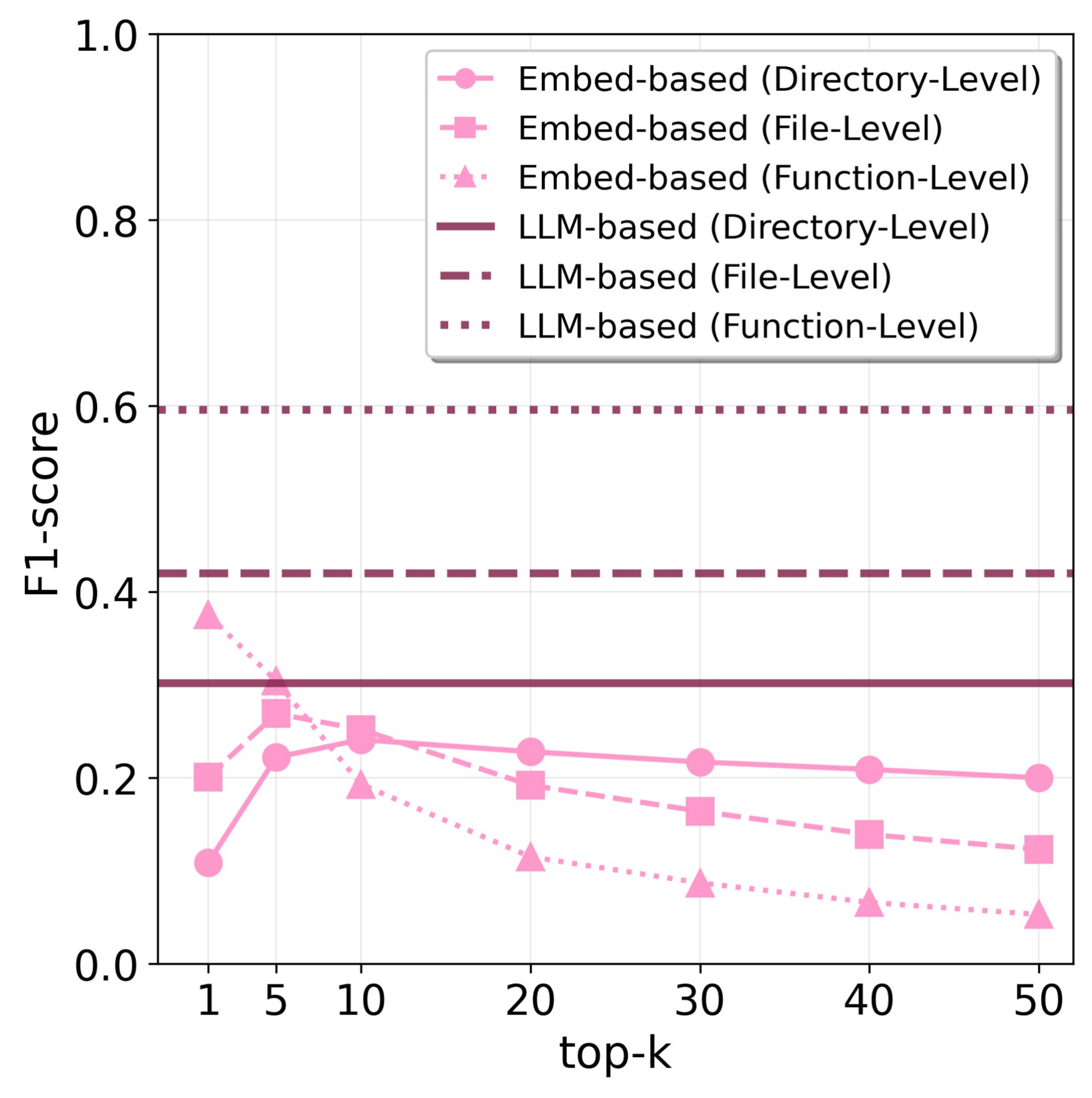}
        \vspace{-6pt}
        \caption{F1 score}
        \label{fig:rq1_f1}
    \end{subfigure}
    \vspace{-6pt}
    \caption{Comparison between LLM-based and embedding-based localization across three metrics}
    \label{fig:rq1_result}
\end{figure*}

\vspace{-2mm}
\section{Preliminary Study}
Before applying our LLM-driven pipeline to declined proposals, we conducted a preliminary study using accepted proposals to compare its performance with the RAG-based TLR method.
We regard this comparison as a preliminary step because prior studies have typically evaluated their TLR methods on accepted proposals, where the discussed code has already been implemented in the repository.
In contrast, our target task involves generating traceability links between declined proposals and source code, which fundamentally differs from the prior setting because the discussed code is not implemented.
Therefore, before addressing this new task, we first verified that our pipeline performs better than the state-of-the-art RAG-based method on accepted proposals to confirm its validity, and then proceeded to the main experiments (RQ1 and RQ2) focusing on declined proposals.


\vspace{-2mm}
\subsection{Approach}
\noindent
\textbf{Data Construction.}
We targeted 337 accepted proposals. For each proposal, we determined an appropriate code granularity (i.e., whether the proposal links to directory-, file-, or function-level code) and constructed ground truth links.
Specifically, since the code granularity is not explicitly defined, the first author inspected each accepted proposal to assign the appropriate granularity based on the criteria developed in \textit{Validation}.
We excluded proposals whose scope targeted the official Go Wiki or subrepositories rather than the main repository. After this filtering, 296 accepted proposals remained.

Next, we extracted the corresponding code changes from Gerrit~\cite{gerrit_history}.
The official Go repository manages code changes and merges through Gerrit, and almost all accepted proposals include URLs referencing the corresponding changes. Once these corresponding changes were identified, we confirmed (1) whether they were labeled as \texttt{MERGED}, and (2) whether their commit messages referred to the corresponding accepted proposal ID (e.g., \texttt{\#12345}).
Because the code changes must have been implemented in the repository, we only targeted accepted proposals linked to \texttt{MERGED} changes. Moreover, our explicit verification of proposal IDs in commit messages helped ensure that the linked changes actually implemented the proposal, thereby reducing false positive links.
As a result, we obtained 262 accepted proposals with ground truth links. The granularity distribution of these 262 proposals is shown in Table~\ref{tab:granularity-dist}.




\begin{table}[t]
  \centering
  \caption{Granularity distribution of proposals}
  \label{tab:granularity-dist}
  \begin{tabular}{lrrrr}
    \toprule
    & Directory & File & Function & Total \\
    \midrule
    Accepted & 17 (6.5\%) & 237 (90.5\%) & 8 (3.1\%) & 262 \\
    Declined & 23 (6.7\%) & 291 (85.3\%) & 27 (7.9\%) & 341 \\
    \bottomrule
  \end{tabular}

  \vspace{2pt}
\end{table}

\noindent
\textbf{Baseline.}
As a baseline, we implemented a RAG-based method following the prior study LiSSA~\cite{fuchss2025lissa}.
Because it was not designed to generate traceability links from proposals to source code, we adapted its pipeline to our task.
The overall framework converts proposal discussions and source code into embedding vectors, computes their similarity to retrieve candidate pairs (\textit{Localization}), and then uses an LLM to determine whether to create links for those retrieved pairs (\textit{Link Decision}).
When constructing the embedding vectors, we followed prior work~\cite{fuchss2025lissa, golzadeh2021identifying, raulji2016stop, jivani2011comparative}, applying standard preprocessing that removes hyperlinks, stop words, and common tags, and performed stemming.



\noindent
\textbf{Evaluation.}
The differences between our pipeline and the RAG-based baseline are: (i) whether Phase~0 is included to determine the appropriate granularity for each proposal, and (ii) whether Phase~1 (\textit{Localization}) uses an embedding model or an LLM.
Since our baseline does not include Phase~0, and Phase~2 is identical in both methods, we executed only the \textit{Localization} phase and compared its performance.

We used standard TLR evaluation metrics~\cite{hayes2006advancing, cleland2012software, fuchss2025lissa}: precision, recall, and F1 score.
LiSSA~\cite{fuchss2025lissa} evaluated performance using the top-20 most similar pairs as candidates; our baseline follows this approach.
We refer to this parameter as top-$k$ and vary it to examine its impact on performance, specifically testing top-1, 5, 10, 20, 30, 40, and 50.




\vspace{-2mm}
\subsection{Results}
Figure~\ref{fig:rq1_result} shows the localization performance of our proposed pipeline and the RAG-based baseline.
The $x$-axis represents the top-$k$ values used in the RAG-based approach.
Because our proposed pipeline does not require a top-$k$ parameter (the LLM directly identifies the most relevant candidates), its performance remains constant across the $x$-axis, appearing as a flat line in the figure.




\textbf{The proposed pipeline achieved higher F1 scores in all cases and higher precision in 21 out of 22 cases compared to the RAG-based approach,} indicating that the LLM-based localization in our pipeline performs more effectively than the embedding-based localization used in the prior RAG-based method.
Although our pipeline was designed to generate traceability links between declined proposals and source code, a task not considered in previous studies, it also outperformed (or at least matched) the state-of-the-art method in generating traceability links between accepted proposals and source code.
This result demonstrates the validity of our design and justifies applying the proposed pipeline to declined proposals in the main experiments, where we evaluate its performance (RQ1) and analyze its failure cases (RQ2).

\vspace{-2mm}
\section{\RQone}
\label{subsec:RQ1}
In this section, we evaluate our proposed pipeline using declined proposals from the Go language repository.

\vspace{-2mm}
\subsection{Approach}
\noindent
\textbf{Data Construction.}
Since the code discussed in declined proposals is not implemented in the repository, it was necessary to identify relevant code for each declined proposal.
As a first step toward this goal, the first author manually determined the appropriate granularity (i.e., directory-, file-, or function-level) for each declined proposal based on the granularity criteria established in the \textit{Validation} phase.
Among the 388 declined proposals remaining after \textit{Initialization} and \textit{Validation}, those whose scope targeted the official Go Wiki or Go subrepositories were excluded, resulting in 341 declined proposals with assigned granularities (Table~\ref{tab:granularity-dist}).

\noindent
\textbf{Evaluation.}
In RQ1, we evaluated the pipeline from two perspectives: whether it selected the appropriate granularity for each proposal (granularity accuracy) and whether it generated correct links at that granularity (precision).
Granularity accuracy was assessed using the ground truth granularities defined in \textit{Data Construction}.

Evaluating all proposal–code pairs to construct a complete set of ground-truth links is infeasible, since it would require manually verifying the relationship between each declined proposal and every directory, file, and function in the entire codebase.
Therefore, as a best-effort approach, the first author manually inspected the links generated by the proposed pipeline and judged their correctness according to the link evaluation criteria developed in \textit{Validation}, from which we calculated precision.
Because not all traceability links could be identified, recall could not be evaluated in this RQ.




Although the evaluation involved manual judgment and was therefore subject to potential subjectivity, we conducted an additional validation to assess its reliability.
The fourth author independently relabeled a randomly sampled subset of 181 out of 341 declined proposals (95\% confidence level, 5\% margin of error).
The results yielded Cohen’s $\kappa$ values of 0.635 for granularity and 0.615 for link correctness, indicating substantial agreement between the two authors~\cite{viera2005understanding}.
These results support the reliability of our manual evaluation in this RQ.


\vspace{-2mm}
\subsection{Results}
Table~\ref{tab:granularity_accuracy_precision} presents the performance of our pipeline in selecting the appropriate granularity (granularity accuracy) and generating correct links (precision).

\textbf{The pipeline achieved an average accuracy of 0.836 in selecting the correct granularity, although there remains room for improvement at the function level.}
As shown in Table~\ref{tab:granularity-dist}, the file-level granularity accuracy was the highest (0.887), leading to the overall average of 0.836.
However, our proposed pipeline struggled to select the correct granularity at the function level, achieving an accuracy of only 0.407.

\textbf{The pipeline generated correct links with an average precision of 0.643, whereas the precision at the function level was significantly lower (0.253).}
The link precision at the file level was the highest (0.688), similar to the trend observed in granularity accuracy.
However, the precision at the function level was only 0.253, which is considerably lower than the directory-level (0.529) and file-level (0.688) values.
This low precision at the function level mainly stems from incorrect granularity selection.
If the pipeline selects the wrong granularity, it cannot generate correct links for that proposal, resulting in a precision of 0.000 for that case.

\summaryblock{Answer to RQ1}{
Our proposed pipeline demonstrated promising performance in linking declined proposals to code, particularly at the directory and file levels.
It achieved an overall granularity accuracy of 0.836 and a precision of 0.643.
However, generating correct links at the finest granularity (function level) remains a challenging task, as the link precision at this level was significantly low (0.253).
}





\begin{table}[t]
\centering
\small
\caption{Granularity accuracy and precision}
\vspace{-6pt}
\label{tab:granularity_accuracy_precision}
\begin{tabular}{lrr}
\toprule
Granularity & Accuracy & Precision \\
\midrule
Directory & 0.696 & 0.529 \\
File      & 0.887 & 0.688 \\
Function  & 0.407 & 0.253 \\
\midrule
Average   & 0.836 & 0.643 \\
\bottomrule
\end{tabular}
\end{table}

\vspace{-2mm}
\section{\RQtwo}
\label{subsec:RQ2}
%

\begin{figure}[t]
    \begin{center}
    \includegraphics[width=.85\linewidth]{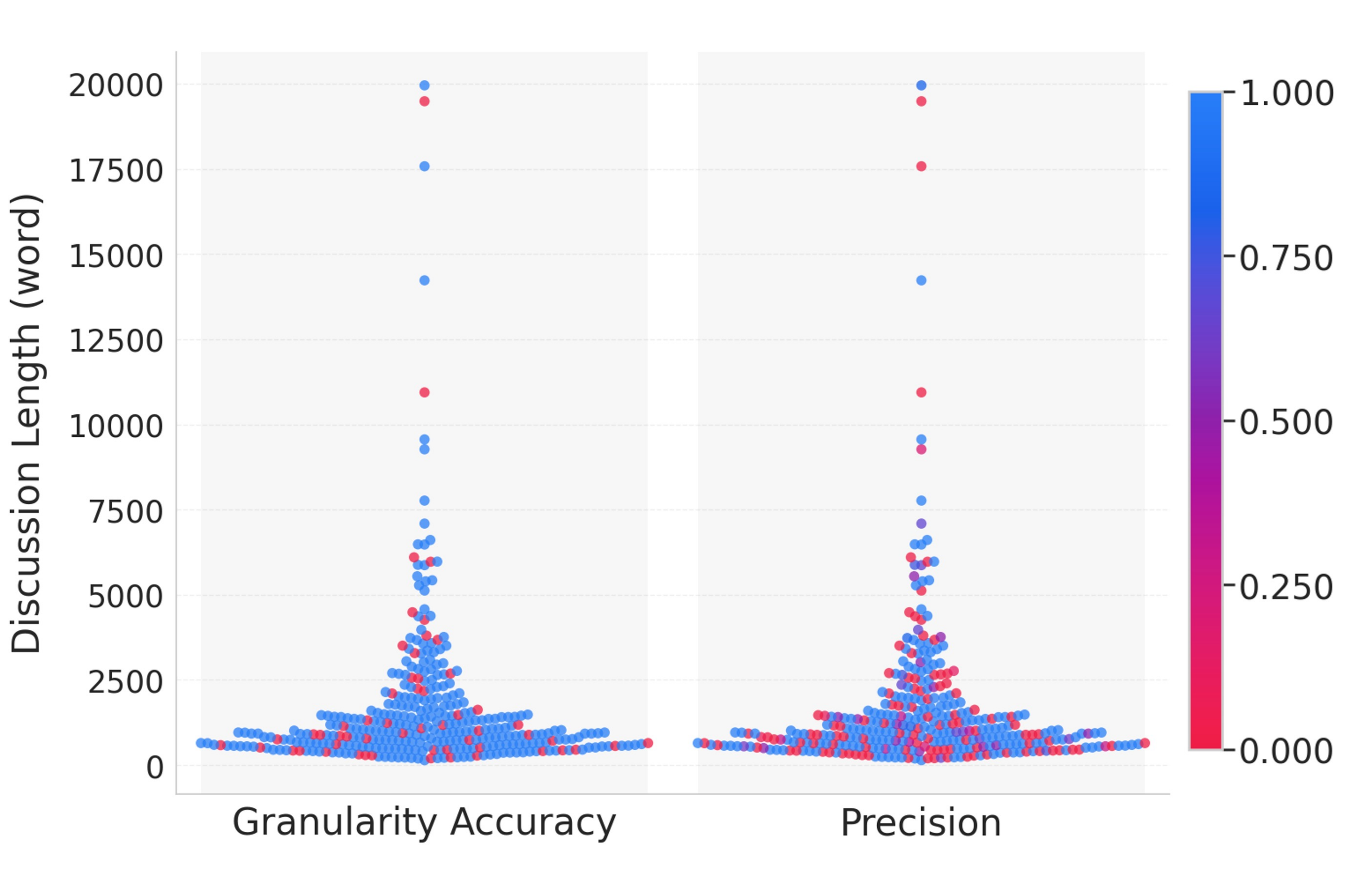}
    \vspace{-6pt}
    \caption{Relationship between performance and discussion length; longest discussion excluded (55083 words, granularity accuracy = 0.000, precision = 0.000 )}
    \label{fig:discussion}
    \end{center}
\end{figure}

In this section, we investigate why our proposed pipeline failed to generate correct links for declined proposals.
One possible factor is that the discussions of declined proposals may be too long, as prior studies~\cite{li2024long, liu2023lost} have reported that LLMs can struggle with in-context learning under long contexts.
However, discussion length did not affect the performance of our pipeline.
Figure~\ref{fig:discussion} visualizes the relationship between discussion length ($y$-axis) and prediction performance (color, where blue indicates higher and red indicates lower performance), showing no apparent correlation.
The Spearman rank correlation further confirmed a small and non-significant association between discussion length and performance: $\rho = 0.035$ ($p = 0.526$).
Therefore, the decline in performance cannot be directly attributed to longer discussions.

\vspace{-2mm}
\subsection{Approach}
\begin{figure}[t]
    \begin{center}
    \includegraphics[width=.65\linewidth]{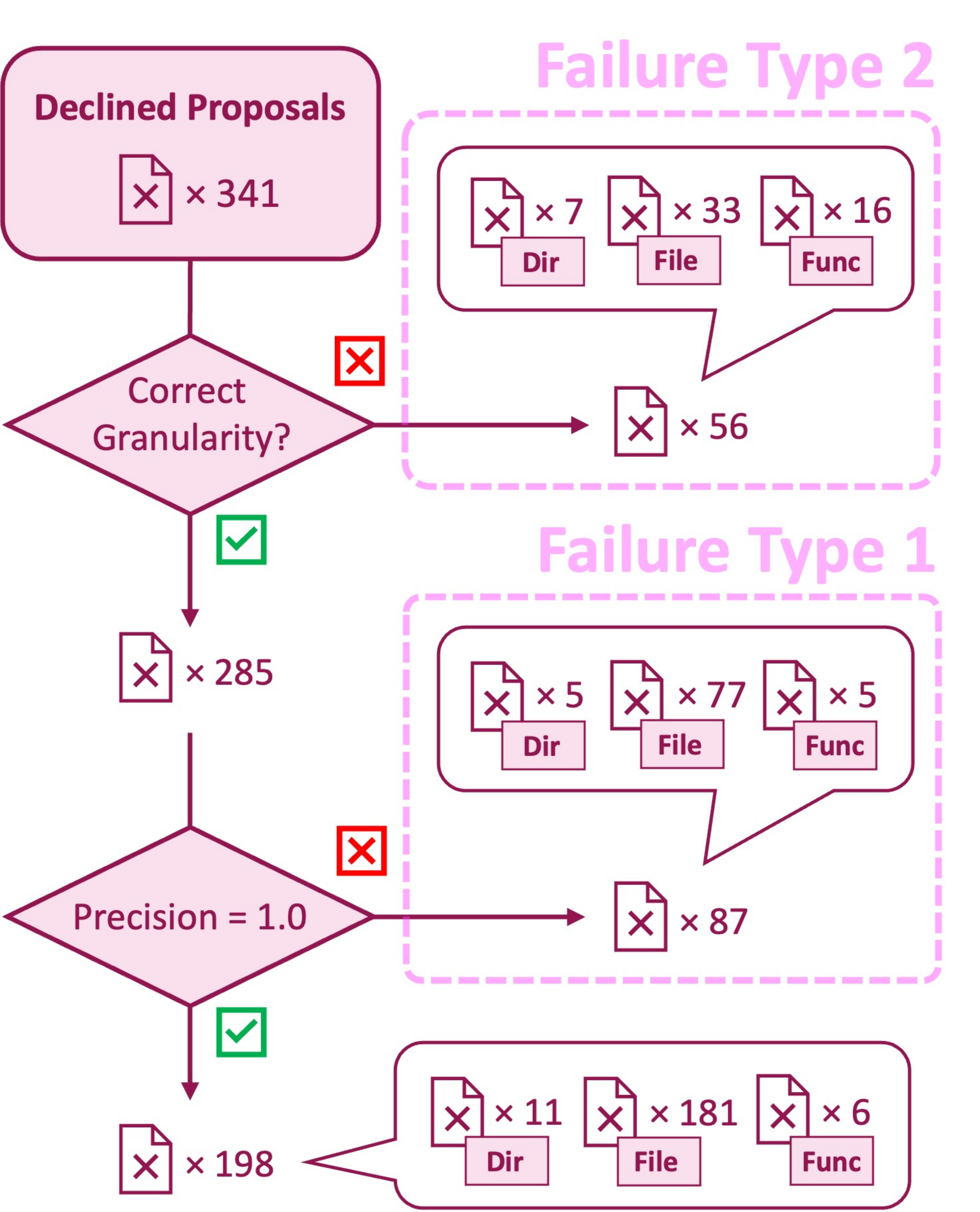}
    \vspace{-6pt}
    \caption{Distribution of failed proposals and scope of the analyses}
    \label{fig:failure_distribution}
    \end{center}
\end{figure}

To explain why the pipeline fails, we investigated two distinctive failure types: (1) cases where the pipeline selected the correct granularity but generated incorrect links, and (2) cases where the pipeline misclassified the granularity.
Figure~\ref{fig:failure_distribution} illustrates these two failure types.
In this RQ, the failure type 1 includes 87 declined proposals out of 285 declined proposals where the pipeline selected the correct granularity; the failure type 2 includes 56 declined proposals out of 341 declined proposals where the pipeline misclassified the granularity.
Based on these two failure types, we cover all failed cases (143 declined proposals).


We performed two complementary analyses:
(A1) a quantitative analysis examining whether the target code appeared in the discussion and how this affected performance, and
(A2) a qualitative analysis through \textit{open coding}.
We first applied (A1) to each of the failure types. When (A1) did not fully explain the failures, we applied (A2). 
Below, we describe each analysis in detail.


\noindent
\textbf{A1: Reference Explicitness $\rightarrow$ Performance.}
This analysis investigates whether the discussion includes explicit mentions of the target code to be linked, and to what extent such mentions affect the performance.
Each proposal was categorized into Level-1 / Level-2 / Level-3. 
We describe these levels below.
\begin{itemize}\setlength{\itemsep}{0pt}
  \item \textit{Level-1 (Direct)}: 
  The proposal explicitly names the target code (file, directory, or function).
  For example, \textit{Add Foo to src/net/http/server.go} directly specifies the linked file.

  \item \textit{Level-2 (Indirect)}:
  The proposal mentions a struct, variable, type, or other element that uniquely identifies the target package or file, even if the file name itself is not stated.
  For example, \textit{Add WriteNopCloser to iouti.NopCloser} implies that the file containing the \textit{NopCloser} struct is the relevant target.

  \item \textit{Level-3 (Implicit)}:
  The proposal does not explicitly indicate the target code location, required contextual inference.
  For example, when it discusses a product file change in directory A and directory A contains only one project file, we can infer that this product file is the relevant target.
\end{itemize}

We then compared the performance across these levels.
We statistically examined the relationship between \textit{Reference Explicitness} and performance using a Kruskal–Wallis test for the failure type 1 and a chi-squared test for the failure type 2.
In this analysis, \textbf{we hypothesized the highest performance at Level-1 and the lowest at Level-3 since explicit mentions should facilitate link generation}.




\noindent
\textbf{A2: Open Coding.}
We applied A2 when our hypothesis in A1 was not supported since it indicates that A1 did not fully explain the failures.
In our analysis, the failure type 2 was not fully explained by A1, so we conducted A2 on these proposals.
We performed an open coding analysis to identify discussion patterns associated with failures and construct a taxonomy of these patterns.
Following prior work~\cite{hata20199}, the analysis proceeded in multiple iterations.
In the first iteration, two researchers independently coded 20 declined proposals.
We then discussed and consolidated a shared coding guide, tested it on another 20 declined proposals, and iteratively refined the coding guide by merging overlapping codes and adding missing ones.
The process concluded after analyzing all 56 declined proposals that are the failure type 2 (Figure~\ref{fig:failure_distribution}).

\vspace{-2mm}
\subsection{Results}

\begin{figure}[t]
    \begin{center}
    \includegraphics[width=.6\linewidth]{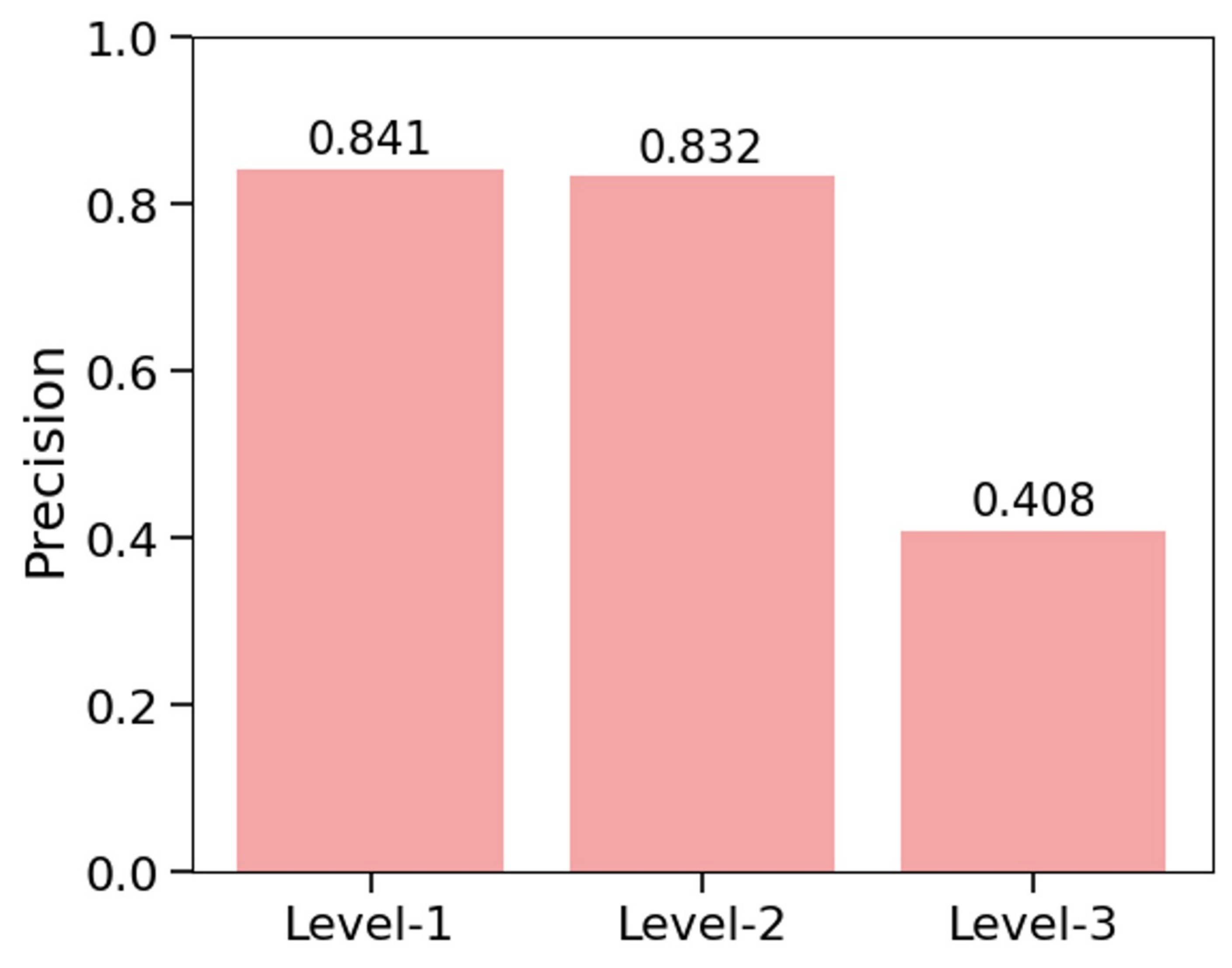}
    \vspace{-6pt}
    \caption{Precision by \textit{Reference Explicitness} for declined proposals in which the pipeline correctly determined the granularity (285 proposals)}
    \label{fig:rq3_precision}
    \end{center}
\end{figure}

\textbf{For the failure type 1, the performance of the pipeline in generating correct links depended on how explicitly the target code was mentioned in the discussion.}
Figure~\ref{fig:rq3_precision} shows the precision for each level of \textit{Reference Explicitness} (Level-1 to 3).
Level-1 and Level-2 show higher precision (0.841 and 0.832), while Level-3 shows lower precision (0.408).
Applying the Kruskal–Wallis test revealed significant differences between Level-1 and Level-2 vs.\ Level-3.
This result indicates that, when the granularity is correctly predicted, the clarity of code references in the discussion strongly contributes to the accuracy of link generation.

\begin{figure}[t]
    \begin{center}
    \includegraphics[width=.6\linewidth]{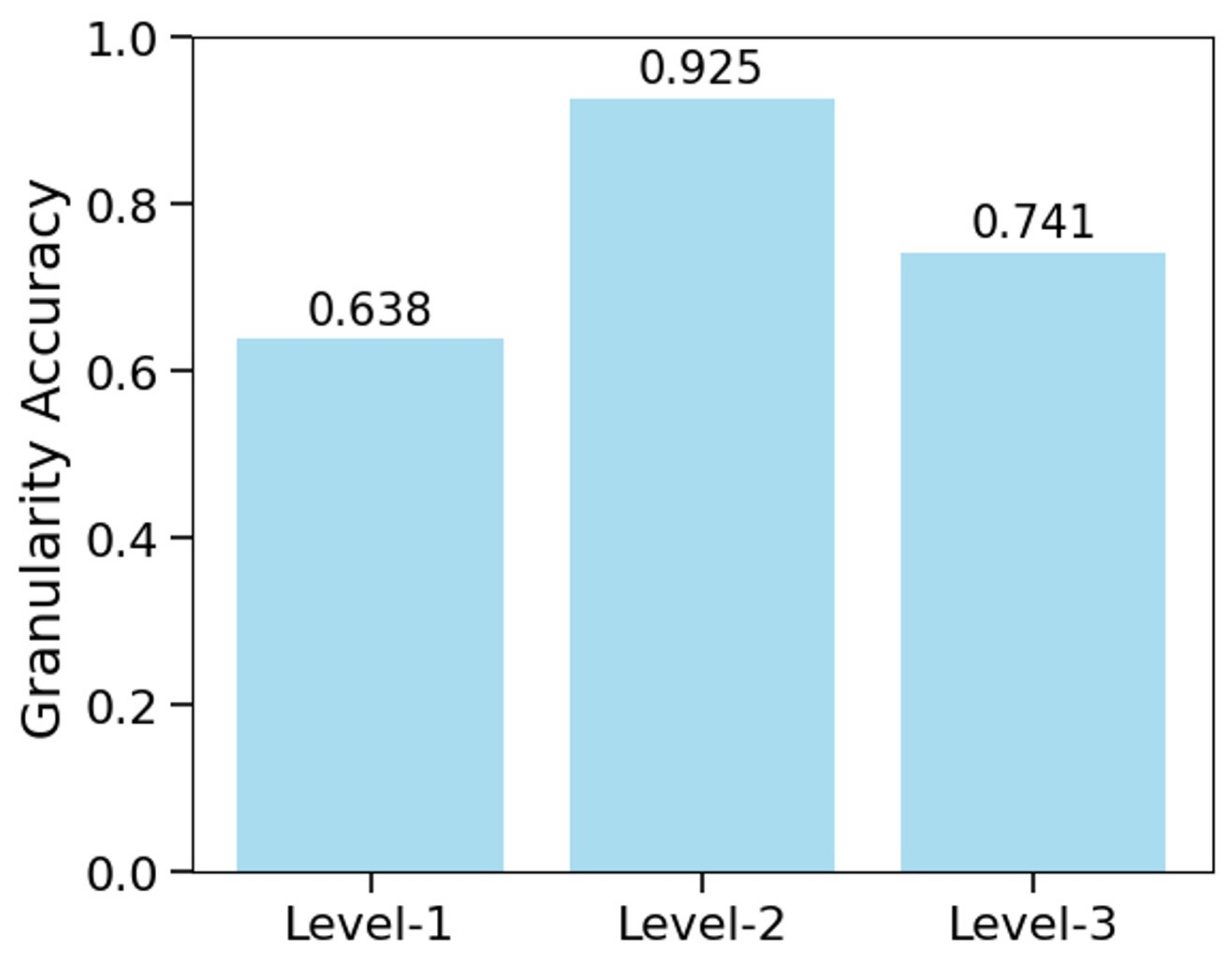}
    \vspace{-6pt}
    \caption{Granularity accuracy by \textit{Reference Explicitness} for all declined proposals (341 proposals)}
    \label{fig:rq3_granularity_accuracy}
    \end{center}
\end{figure}

\textbf{For the failure type 2, about half of the proposals lacked concrete guidance on how to implement the change.}
Figure~\ref{fig:rq3_granularity_accuracy} shows the granularity accuracy for each level of \textit{Reference Explicitness} (Level-1 to 3).
This figure clearly shows that our hypothesis is not supported since Level-2 shows the highest granularity accuracy (0.925), followed by Level-3 (0.741) and Level-3 (0.638).
Since this result suggests that A1 did not fully explain the failures in this type, we conducted A2 on these proposals.

Table~\ref{tab:granularity_error_codes} presents the taxonomy of discussion patterns identified through our open coding.
Below, we describe each of the categories.
\begin{itemize}\setlength{\itemsep}{0pt}
\item \textit{Insufficient Implementation Guidance}:
The discussion describes what to implement but does not specify how or where to implement it.
\item \textit{Conflicting Implementation Suggestions}:
The proposal simultaneously presents multiple or inconsistent implementation approaches.
\item \textit{Hidden Implementation Guidance}:
Key indicators for granularity decisions are hidden in lengthy discussions, making them difficult to identify.
\item \textit{Confusing Granularity References}:
The discussion includes lexical or contextual references; however, they refer to a different granularity level.
\item \textit{Unknown Reason}:
The discussion appears to include sufficient information for a granularity decision, yet the reason for misclassification remains unclear.
\end{itemize}
A large proportion (42.9\%) of the proposals were labeled as \textit{Insufficient Implementation Guidance}, indicating that essential implementation information is often missing when the pipeline fails to choose the appropriate granularity.
\textit{Conflicting Implementation Suggestions}, \textit{Hidden Implementation Guidance}, and \textit{Confusing Granularity References} indicate that, although the necessary information is present, other content may obscure or distort the granularity decision.





These analyses show that failures to generate correct links fall into two categories:
(1) the discussion lacks the information required to make the linking decision, and
(2) the required information is present but obscured or confounded by extraneous content.
To improve performance, both issues must be addressed.
When information is missing, supplementary artifacts referenced in the discussion (e.g., related proposals or code elements) should be curated and provided as additional input to the pipeline.
When key signals are obscured, a summary that retains only the information necessary for linking should be generated and used as the pipeline input.
These two directions highlight important avenues for future research on generating traceability links between declined proposals and source code.




\begin{table}[t]
  \centering
  \caption{Granularity-error patterns}
  \vspace{-6pt}
  \label{tab:granularity_error_codes}
  \footnotesize
  \begin{tabular}{lrr}
    \toprule
    \textbf{Code} & \textbf{Count} & \textbf{Pro (\%)} \\
    \midrule
    Insufficient Implementation Guidance & 23 & 42.9 \\
    Conflicting Implementation Suggestions & 14 & 23.2\\
    Hidden Implementation Guidance & 6 &  10.7\\
    Confusing Granularity References & 5 & 8.9\\
    Unknown Reason & 8 & 14.3\\
    \bottomrule
  \end{tabular}
\end{table}

\summaryblock{Answer to RQ2}{
Our proposed pipeline fails to identify the correct granularity because of (1) missing or insufficient implementation details in the discussion and (2) extraneous content that obscures key information.
When granularity is correct, link accuracy depends on how explicitly the target code is mentioned--clear code references lead to higher precision.
%
}

\section{Threats to Validity}
We disclose the threats to validity and our mitigation strategies. 

\noindent
\textbf{Construct Validity.}
Since the code discussed in declined proposals is not implemented in the repository, the traceability links between declined proposals and their corresponding code cannot be automatically detected from development history.
To establish traceability links, each declined proposal would need to be manually examined to determine whether it is linked to any code artifacts (directory, file, or function), which is extremely time-consuming.
Therefore, we primarily evaluated precision in this study (RQ1); recall could not be assessed under this setting.

Moreover, our evaluation relied on manual assessments by the authors.
To mitigate subjectivity, we developed and published coding guides. 
Also, we conducted these assessments by multiple authors and measured inter-rater agreement.\\

\noindent
\textbf{Internal Validity.}
The outputs of the pipeline depend on the behavior of the underlying LLM.
To mitigate variability, we set the temperature to 0.0; nevertheless, the results remain sensitive to prompt design and how proposal discussions and repository context are provided as inputs to the model.
Although our objective in this study is to formalize the task of linking declined proposals to source code rather than to propose the state-of-the-art TLR method, the pipeline design still leaves room for improvement through alternative models, prompt strategies, and tuning parameters.
To address these issues and facilitate reproducibility and extension, we provide a comprehensive replication package containing all data, prompts, and scripts used in this study.
This package allows future researchers to easily reproduce our experiments, explore the impact of different LLM configurations, and empirically extend the pipeline with improved models or prompt techniques.\\




\noindent
\textbf{External Validity.}
Our study used proposals from the official Go language repository.
The proposal process of Go is rigorous and well documented, making it an appropriate subject for this study.
In particular, the process behind decisions to accept or decline proposals is clearly defined (e.g., developers are required to actively discuss a proposal before a decision is made), which enables this study to infer relevant information about the source code even when the proposals do not explicitly mention it.
However, since the evaluation focused solely on the Go project, future work should expand the scope to other software systems to improve generalizability, as the concept of declined proposals is not unique to Go and can be extended to other open-source projects.
For example, the PEP (Python Enhancement Proposal) process provides structured artifacts, including concise technical specifications and accompanying rationales, which make it a promising target for extending our study to another language ecosystem~\cite{pep1}.

\vspace{-2mm}
\section{Conclusion \& Future Work}
\label{sec:summary}
We propose a novel linking task between declined proposals and source code and proposed an LLM-based pipeline to address this task.
Our evaluation shows that (1) the proposed pipeline achieves higher F1 scores and precision than the state-of-the-art RAG-based baseline on accepted proposals, and (2) for declined proposals, it can select the correct granularity with 0.836 accuracy and generate correct links with a mean precision of 0.643.
In addition, our failure analysis indicates that the performance of our pipeline depends more on the content of the discussion than on its length.
Specifically, the pipeline struggles (1) when the discussion lacks sufficient implementation details and (2) when it contains irrelevant information that obscures important details.
When the pipeline successfully identifies the correct granularity, its performance heavily depends on how explicitly the target code is mentioned in the discussion.

Based on these findings, we suggest the following two directions for future research on linking declined proposals to source code:
\begin{itemize}\setlength{\itemsep}{0pt}
\item When information in the discussion is missing or insufficient, augmenting it with relevant context from external sources such as related proposals or code comments could help improve performance.
Future work should explore such augmentation techniques to enhance the quality of inputs for LLMs.
\item When important details are obscured by irrelevant content, summarizing the discussion to highlight key points could help LLMs focus on relevant information.
Future research should investigate summarization techniques to improve the clarity and usefulness of the input provided to LLMs.
\end{itemize}
\vspace{-2mm}
\section{Data Availability}
Our replication package can be accessed at \url{https://doi.org/10.5281/zenodo.18732456}.

\section*{Acknowledgement}
We gratefully acknowledge the financial support of: (1) JSPS for the KAKENHI grants (JP25K03100);
(2) Japan Science and Technology Agency (JST) as part of Adopting Sustainable Partnerships for Innovative Research Ecosystem (ASPIRE), Grant Number JPMJAP2415,
and (3) the Inamori Research Institute for Science for supporting Yasutaka Kamei via the InaRIS Fellowship.

\balance
\bibliographystyle{ACM-Reference-Format}
\bibliography{references}

\end{document}